\documentclass{IOS-Book-Article}

\usepackage{mathptmx}
\usepackage{graphicx}
\usepackage{multirow}
\usepackage{enumitem}
\usepackage{xcolor}
\usepackage{float}
\usepackage{url}
\usepackage[colorlinks]{hyperref}

\newcommand{\eidle}{$\mathbf{E_{\texttt{idle}}}$}
\newcommand{\efull}{$\mathbf{E_{\texttt{full}}}$}

\newcommand{\ework}{$\mathbf{E_{\texttt{work}}}$}
\newcommand{\eworkcu}{$\mathbf{E_{\texttt{work-CU}}}$}
\newcommand{\tworkcu}{$\mathbf{T_{\texttt{work-CU}}}$}
\newcommand{\ourcode}{\sc{Hy-Nbody}}
\newcommand{\globalm}{\textit{global}}
\newcommand{\localm}{\textit{local}}
\newcommand{\CU}{\texttt{CUs}}

\def\hb{\hbox to 10.7 cm{}}

\begin{document}

\pagestyle{headings}
\def\thepage{}

\begin{frontmatter}              

\title{Performance and energy footprint assessment of FPGAs and GPUs on HPC systems using Astrophysics application}

\markboth{}{March 2020\hb}

\author[A]{\fnms{David} \snm{Goz}%
\thanks{Corresponding Author: David Goz, ORCID: 0000-0001-9808-2283, INAF-Osservatorio Astronomico Di Trieste,
Via G.B. Tiepolo 11, 34131 Trieste, Italy;
E-mail:
david.goz@inaf.it}},
\author[B]{\fnms{Georgios} \snm{Ieronymakis}}
,
\author[B]{\fnms{Vassilis} \snm{Papaefstathiou}}
,
\author[B]{\fnms{Nikolaos} \snm{Dimou}}
,
\author[A]{\fnms{Sara} \snm{Bertocco}}
,
\author[A]{\fnms{Igor} \snm{Coretti}}
,
\author[C]{\fnms{Francesco} \snm{Simula}}
,
\author[A]{\fnms{Antonio} \snm{Ragagnin}}
,
\author[A]{\fnms{Luca} \snm{Tornatore}}
and
\author[A]{\fnms{Giuliano} \snm{Taffoni}}

\runningauthor{D. Goz et al.}
\address[A]{INAF-Osservatorio Astronomico di Trieste, Italy}
\address[B]{FORTH-ICS, Heraklion, Crete, Greece}
\address[C]{INFN-Sezione di Roma, Italy}

\begin{abstract}

New challenges in Astronomy and Astrophysics (AA) are urging the need for a large number of exceptionally computationally intensive simulations. "Exascale" (and beyond) computational facilities are mandatory to address the size of theoretical problems and data coming from the new generation of observational facilities in AA.
Currently, the High Performance Computing (HPC) sector is undergoing a profound phase of innovation, in which the primary challenge to the achievement of the ``Exascale'' is the power-consumption. 
The goal of this work is to give some insights about performance and energy footprint of contemporary architectures for a real astrophysical application in an HPC context.
We use a state-of-the-art N-body application that we re-engineered and optimized to exploit the heterogeneous underlying hardware fully. We quantitatively evaluate the impact of computation on energy consumption when running on four different platforms. Two of them represent the current HPC systems (Intel-based and equipped with NVIDIA GPUs), one is a micro-cluster based on ARM-MPSoC, and one is a "prototype towards Exascale" equipped with ARM-MPSoCs tightly coupled with FPGAs. We investigate the behaviour of the different devices where the high-end GPUs excel in terms of time-to-solution while MPSoC-FPGA systems outperform GPUs in power consumption.
Our experience reveals that considering FPGAs for computationally intensive application seems very promising, as their performance is improving to meet the requirements of scientific applications.
This work can be a reference for future platforms development for astrophysics applications where computationally intensive calculations are required.

\end{abstract}

\end{frontmatter}
\markboth{February 2019\hb}{February 2019\hb}

\section{Introduction and motivation}
\label{section:intro}

In the last decade, energy efficiency has become the primary concern in the High Performance Computing (HPC) sector. HPC systems constructed from conventional multicore Central Processing Units (CPUs) have to face, on one side, the reduction in year-on-year performance gain for CPUs and on the other side, the increasing cost of cooling and power supply as HPC clusters grow larger.

Some technological solutions have already been identified to address the energy issue in HPC \cite{dutot}; one of them is the use of power-efficient Multiprocessor Systems-on-Chip (MPSoC) \cite{cesiniCosa,Brain_LowPower,Power_DPSNN,Calore_2015,Nikolskiy_2016,Morganti_2016,Taffoni_ARM}. These hardware platforms are integrated circuits composed of multicore CPUs combined with accelerators like Graphic-Processing-Units (GPUs) and/or Field-Programmable-Gate-Arrays (FPGAs). Such hardware accelerators can offer higher throughput and energy-efficiency compared to traditional multicore CPUs. The main drawback of those platforms is the complexity of their programming model, requiring a new set of skills for software developers and hardware design concepts, leading to increased development time for accelerated applications.

The Astronomy and Astrophysics (AA) sector is one of the research areas in physics that requires more and higher performing software, as well as the necessity of Exascale supercomputers (and beyond) \cite{taffoni}. In AA, HPC numerical simulations are the most effective instruments to model complex dynamic systems, to interpret observations and to make theoretical predictions, advancing scientific knowledge. They are mandatory to help capture and analyze the torrent of complex observational data that the new generation of observatories produce, providing new insights into astronomical phenomena, the formation and evolution of the universe, and the fundamental laws of physics. 

The research presented in this paper arises in the framework of European Exascale System Interconnect and Storage (EuroExa) European funded project\footnote{EuroExa: \url{https://euroexa.eu/}} aiming at the design and development of a prototype of an exascale HPC machine. EuroEXA is achieving that through the use of low-power ARM processors accelerated by tightly-coupled FPGAs. 

Building an Exascalable platform, i.e. a supercomputer able to reach a peak performance greater than $10^{18}$ FLoating point Operations Per Second (FLOPS), is a clear priority that is spurring worldwide initiatives. 
All the different proposed approaches imply the re-design of the underlying technologies (i.e., processors, interconnect, storage, and accelerators) both to improve their performances and to reduce their energy requirements by about one order of magnitude. On the other side, Exascale platforms are composed by an extremely large number of nodes (e.g. in the case of EuroEXA platform we expect $10^6$ nodes) "glued" together by a low latency and high throughput interconnect. This implies that a large effort also on usability (e.g. system software), reliability and resiliency must be conceived. 
Efficiently running scientific applications therein requires to exploit not only CPUs and accelerators, but also the complex memory hierarchy supported by the multi-hierarchy interconnect available on the platforms. 

Focusing on performance and energy-efficiency, in this work we exploit four platforms:
\begin{itemize} 
\item (I-II) two Linux x86 HPC clusters that represent the state-of-the-art of HPC architectures (Intel-based and equipped with NVIDIA GPUs); 
\item (III) a Multiprocessor SoC micro-cluster that represents a low purchase-cost and low-power approach to HPC; 
\item (IV) an exascale prototype that represents a possible future for supercomputers. This prototype was developed by the  ExaNeSt European project\footnote{ExaNeSt: \url{https://exanest.eu/}} \cite{exanest, exanest2, KATEVENIS201858} and customized by the EuroEXA project.
\end{itemize}

The platforms are probed using a direct $N$-body solver for astrophysical simulations, widely used for scientific production in AA, e.g. for simulations of star clusters up to $\sim 8$ million bodies \cite{higpus_1, higpus_2}.
Even if, the role of the interconnect is one of the core issued in designing an exascale platform, the application studied in this paper is a CPU bound code, so different interconnect technologies implemented by the 4 platforms do not affect the results discussed.

The goal of this paper is to investigate the performance-consumption plane, namely the parameter space where time-to-solution and energy-to-solution are combined, exploiting the different devices hosted on the platforms. We include the comparison among high-end CPUs, GPUs and MPSoCs tightly coupled with FPGAs systems. To the best of our knowledge, this paper provides one of the first comprehensive evaluations of a real AA application on an exascale prototype, comparing the results with today’s HPC hardware.

The paper is organized as follows. In Section~\ref{section:infrastructures} we describe the computing platforms used for the analysis. In Section~\ref{section:methodology} and ~\ref{section:power_measurements} we discuss the methodology employed to make the performance and energy measurement experiments, including considerations on the usage of the different platforms and the configuration of the parallel runs. Section~\ref{section:code} is devoted to present the scientific application used to benchmark the platforms. Our results are presented in Section~\ref{section:results}. The final Section~\ref{section:conclusion} is devoted to the conclusions and the perspectives for future work.

\section{Computing platforms}
\label{section:infrastructures}
In this section, we describe the four platforms used in our tests.
In Table~\ref{table:platforms}, we list the devices, and we highlight in bold the ones exploited in this paper.

\begin{table*}[h]
\caption{The computing node and the associated devices.
The devices exploited are highlighted in bold.}
\label{table:platforms}
\begin{center}
\begin{tabular}{|c|c|c|c|}

\hline
\textbf{Node}  & \textbf{CPU}                           & \textbf{GPU}                    & \textbf{FPGA} \\
\hline
mC             & 4x(\textbf{ARM A53}) + 2x(ARM A72)             & \textbf{ARM Mali-T864}          & None \\
\hline
IC             & 40x(\textbf{Xeon Haswell E5-4627v3})    & None                            & None \\
\hline
ExaBed         & 16x(ARM A53) + 8x(ARM R5)                       & 4x(ARM Mali-400)                    & 4x(\textbf{Zynq-US+})\\
\hline
GPUC           & 32x(Xeon Gold 6130)                     & 8x(\textbf{Tesla-V100-SXM2})        & None\\
\hline

\end{tabular}
\end{center}
\end{table*}

\subsection{Intel cluster}

Each node of the Intel cluster (hereafter IC) is equipped with 4 sockets INTEL Haswell E5-4627v3 at 2.60 GHz with 10 cores each and 256 GB (6 GB per core).
The theoretical peak performance of each core is 83.2/41.6 GFLOPS for FP32/FP64 respectively.
The interconnect is the Infiniband ConnectX-3 Pro Dual QSFP+ (54Gbs), and the storage system is a BeeGFS parallel  file  system, with 4 IO servers offering 350TB of disk space \cite{bertocco,taffoni_chipp}. 
Each computing node is equipped with an iLO4 management controller, that can be used to measure the node instantaneous power consumption ($1$ sample every second).

\subsection{GPU cluster}

Each node of the GPU cluster (hereafter GPUC) is equipped with 2 sockets INTEL Xeon Gold 6130 at 2.10 GHz with 16 cores each along with 8 NVIDIA Tesla-V100-SXM2. 
The theoretical peak performance of each GPU is 15.67/7.8 TFLOPS for FP32/FP64 respectively.
The GPUs are hosted by a SuperServer 4029GP-TVRT system by SuperMicro\textsuperscript{\textregistered}, which integrates a Baseboard Management Controller (BMC) that through Intelligent Platform Management Interface (IPMI) provides \mbox{out-of-band} access to the sensors embedded into the system.
Among the physical parameters that these sensors are able to measure (temperature, cooling fans speed, chassis intrusion, etc.), this system is also able to continuously monitor amperage and voltage of the different rails within the redundant power supply units, in order to give at least a ballpark figure of its wattage.
Right after booting and with all GPUs in idle, the system wattage is given at about 440~Watts. In order to get full throttle GPU's power measurements we rely on the built-in sensors queried by NVIDIA \texttt{nvidia-smi} tool\footnote{The readings are accurate to within +/- 5 Watts, as stated by NVIDIA documentation. That accuracy does not affect our results.}.

\subsection{ARM-Micro-Cluster}

We design our ARM-Micro-Cluster (hereafter mC) starting from the OpenSource MPSoC Firefly-RK3399 \cite{incass}. This single-board is equipped with the big.LITTLE architecture: 4x(Cortex-A53) cores with 32kB L1 cache and 512kB L2 cache, and a cluster of 2x(Cortex-A72) high-performance cores with 32kB L1 cache and 1M L2 cache. Each cluster operates at independent frequencies, ranging from 200MHz up to 1.4GHz for the LITTLE and up to 1.8GHz for the big. The MPSoC contains 4GB DDR3 - 1333MHz RAM. The MPSoC features also the OpenCL-compliant Mali-T864 embedded GPU that operates at 800 MHz. The theoretical peak performance of each A53 core is 11.2/2.8 GFLOPS for FP32/FP64, and of each A72 core is 14.4/3.6 GFLOPS for FP32/FP64 respectively. The theoretical peak performance of the T-864 GPU is 109/32 GFLOPS for FP32/FP64 respectively.
The ARM-Micro-Cluster, composed by 8 Firefly-RK3399 single-boards, is based on Ubuntu 18.04 Linux and scheduled using SLURM \cite{slurm}. The interconnect is based on Gigabit Ethernet, and the storage system is a device shared via NFS.

\subsection{ExaNest HPC testbed prototype}

\label{section:exascale_proto}

\begin{figure}
    \centering
    \includegraphics[width=\textwidth]{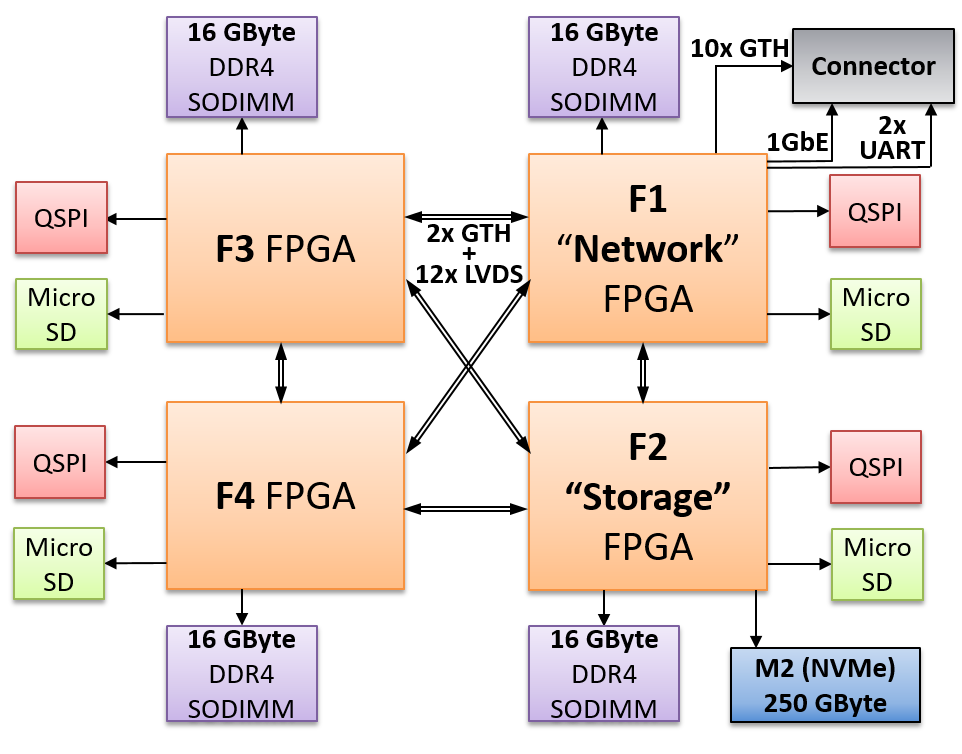}
    \caption{The Quad-FPGA daughterboard block diagram and interconnects.}
    \label{fig:qfdbdiagram}
\end{figure}

The ExaNest HPC testbed prototype \cite{exanest} (hereafter ExaBed) is a liquid-cooled cluster composed of the proprietary Quad-FPGA daughterboard (QFDB) \cite{qfdb} computing nodes, interconnected with a custom network and equipped with a BeeGFS parallel filesystem. In Figure~\ref{fig:qfdbdiagram}, we present a block diagram of the computing node of the platform.

The compute-node board includes 4 Xilinx Zynq Ultrascale+ MPSoC devices (ZCU9EG), each featuring 4x(ARM-A53) and 2x(ARM-R5) cores, along with a rich set of hard IPs and  Reconfigurable Logic. Each Zynq device has a 16GB DDR4 (SODIMM) attached and a 32MB Flash (QSPI) memory. Also, as shown in Figure~\ref{fig:qfdbdiagram}, within the QFDB each FPGA is connected to each other through 2 HSSL and 24 LVDS pairs (12 in each direction). Out of the four, only the "Network" FPGA is directly connected to the outside world, while the "Storage" FPGA has an additional 250 GB M.2 SSD attached to it.
The maximum sustained power of the board is 120 Watts, while the power dissipation during normal operation is usually around to 50 Watts. Targeting a compact design, the dimension of the board is 120-130mm while no component on top or below the printed circuit board (PCB) is taller than 10mm.

These compute nodes are sealed within a blade enclosure, each hosting 4 QFDBs. Currently, the ExaNest prototype HPC testbed consists of 12 fully functional blades. The rack provides connectivity between the blades, while each QFDB is managed through a Manager VM and runs a customized version of Linux based on Gentoo Linux, which is called Carvoonix.

In the QFDB, the measurement of the current and power is accomplished by using a set of TI INA226 coupled with high-power shunt resistors.
The INA226 minimal capture time is 140[$\nu s$]. However, the Linux driver default (and the power-on set-up) sets capture time to 1.1[ms]. The Linux driver also enables averaging from 16 samples, and captures both the shunt and the bus voltages. 
To collect data from the sensors, each board includes 15 I2C power sensors, which allow the measurement of power consumption by major subsystems.

\section{Methodology and considerations}
\label{section:methodology}
The platforms exhibit different behaviour as concern the power policies.

In the case of ARM sockets, the frequency scaling is absent, meaning that \texttt{idle} and \texttt{performance} mode are mutually-exclusive active. Our code, described in the Section~\ref{section:code}, is not able to exploit the highly heterogeneous big.LITTLE ARM socket, whose architecture couples relatively power-saving and slower processor cores (LITTLE) with relatively more powerful and power-hungry ones (big). This MPSoC is conceived to migrate more demanding threads on the more powerful cores of the big socket (A72 in the case of the mC).
Hence, in order to disentangle the performance and the power consumption, we pin all MPI (Message Passing Interface) processes and Open Multi-Processing (OpenMP) threads to the LITTLE socket setting explicitly CPU affinity. It is worth to be noticed that both mC and ExaBed are equipped with A53x4/socket, letting us to run our simulations only on the former and extrapolating the results using CPUs for the latter as well. Hence, we consider it useful to focus on the performance of the FPGA in the Xilinx Zynq UltraScale+ MPSoC hosted by ExaBed, which is in turn the most important topic of this paper. 

On the IC, the ExaBed, the mC and the GPUC respectively, the smallest units for which energy consumption can be measured are a 4-sockets (with 10 cores each) node, a QFDB (4 MPSoCs, with 4 cores and one FPGA each), a single-board (dual socket, with 4 and 2 cores respectively, and one gpu), and one GPU. To carry out a meaningful comparison, we decide to perform the comparison using the same amount of computational units. 
Given the heterogeneity of the platforms in terms of the underlying devices (Table~\ref{table:platforms} as reference), we define the \texttt{computational unit} as a group of four cores for CPU, and either one GPU or FPGA for accelerators. We fix at four cores the \texttt{computational unit} for CPU because both mC and ExaBed are equipped with A53x4 socket.
Table~\ref{table:compute_units} summarizes the \texttt{compute units} (hereafter {\CU}), as defined above, available for each platform.

\begin{table*}[h]
\caption{The \texttt{compute units}, as defined in Section~\ref{section:methodology}, available on the platforms.}
\label{table:compute_units}
\begin{center}
\begin{tabular}{|c|c|c|c|c|}
\hline
\multicolumn{5}{|c|}{{\CU}}                                                          \\ \hline
\textbf{}        & \multicolumn{4}{c|}{\textbf{Platform}}                                                          \\ \hline
\textbf{CU-type} & \textit{\textbf{IC}} & \textit{\textbf{mC}} & \textit{\textbf{ExaBed}} & \textit{\textbf{GPUC}} \\ \hline
\textbf{CPU}     & 10     & 1              & None                     & None                   \\ \hline
\textbf{GPU}     & None                 & 1        & None                     & 8    \\ \hline
\textbf{FPGA}    & None                 & None                 & 4            & None                   \\ \hline
\end{tabular}
\end{center}
\end{table*}

One of the aims of this work is to shed some light on the crucial comparison between the energy consumption and performance of current platforms and (possibly) exascale-like ones.

\section{Power consumption measurements}
\label{section:power_measurements}
Since the IC, Exabed and GPUC have built-in sensors, for those three platforms, we rely on the power measurements returned by the diagnostic infrastructure.
The mC, on the contrary, does not have any sensor, so we obtain the energy consumption by measuring the actual absorption using a Yokogawa WT310E Digital Power Meter.

We assess that having different methods of energy measurements is not affecting the results. To make power measurements, we set-up simulations so that their runtime is much larger than the sampling time of on-board sensors so that fluctuations are averaged out.

\begin{table*}[h]
\caption{The energy consumption of platforms.
{\eidle} is the average power over 3 minutes in idle of the platform; {\efull} is the average energy used over 3 minutes of {\ourcode} continuous execution with 100\% load, using \textbf{one {\texttt{CU}}} of type either CPU, GPU, or FPGA ($\mathbf{E_{\texttt{full-CPU}}}$, $\mathbf{E_{\texttt{full-GPU}}}$, and $\mathbf{E_{\texttt{full-FPGA}}}$ respectively).}
\label{table:platform_consumption}
\begin{center}
\begin{tabular}{|l|c|c|c|c|}
\hline
\multirow{2}{*}{$\mathbf{E[W]}$} & \multicolumn{4}{c|}{Platforms} \\ \cline{2-5} 

\multicolumn{1}{|c|}{}                 & \textbf{\textit{IC}}  & \textbf{\textit{mC}}           & \textbf{\textit{ExaBed}}  & \textbf{\textit{GPUC}} \\ \hline
$\mathbf{E_{\texttt{idle}}}$           & 160                   & 3.15                           & 42.5   & 440   \\ \hline
$\mathbf{E_{\texttt{full-CPU}}}$  & 223                   & 4.55     & N/A    &  N/A  \\ \hline
$\mathbf{E_{\texttt{full-GPU}}}$ & N/A                   & 4.75                           & N/A    &   710 \\ \hline
$\mathbf{E_{\texttt{full-FPGA}}}$ & N/A                   & N/A                            & 53.5   & N/A \\ \hline
\end{tabular}
\end{center}
\end{table*}

For each platform, we measure both the energy consumed under no workload ({\eidle}) and the total energy consumption under $100\%$ workload ({\efull}) using the {\CU}. In Table~\ref{table:platform_consumption} the {\eidle} and {\efull} using one \texttt{compute unit} (of different type) for each platform are shown.

In the following we report the energy-to-solution (total energy required to perform the calculation) excluding the \eidle, i.e $\mathbf{E_{\texttt{work-CU}}} = \mathbf{E_{\texttt{full-CU}}} - \mathbf{E_{\texttt{idle}}}$, in order to focus on the power consumption of different {\CU}.
The idle energy and the energy consumed by the processing units (CPUs, GPUs, FPGAs) are distinct targets for engineering and improvement, and it seems useful to disentangle them while considering what is most promising in the Exascale perspective.

Finally, we estimate the energy impact of the application also in terms of Energy Delay Product (EDP).
The EDP proposed by Cameron \cite{edp} is a "fused" metric to evaluate the trade-off between time-to-solution and energy-to-solution. It is defined as:
\begin{equation}
    \mathbf{EDP_{\texttt{CU}} =  E_{\texttt{work-CU}} \times T_{\texttt{work-CU}}^{w}}
\end{equation}
where {\eworkcu} is the {\ework} consumed during the run by the \texttt{CU}, {\tworkcu} is the time-to-solution of the given \texttt{CU} and $\mathbf{\texttt{w}}$ (usually \texttt{w}=1,2,3) is a parameter to weight
performance versus power. The larger is $\mathbf{\texttt{w}}$ the greater the weight we assign to its performance.

\section{Astrophysical code}
\label{section:code}
As aforementioned, we compare both time-to-solution and energy-to-solution performance using a real scientific application coming from the astrophysical domain: the {\ourcode} code \cite{mynbody,mynbody2}.

In Astrophysics the $N$-body problem consists of predicting the individual motion of celestial bodies interacting purely gravitationally. Since every body interacts with all the others, the computational cost scales as $O(N^{2})$, where $N$ is the number of bodies.
{\ourcode} is a modified version of a GPU-based $N$-body code \cite{nbody:4,nbody:5,nbody:6}, it has been developed in the framework of the ExaNeSt project \cite{exanest}, and it is currently optimized for exascale-like machines within the FET HPC H2020 EuroEXA project.
The code relies on the 6th order Hermite integration schema \cite{hermite}, which consists of three stages: a \texttt{predictor} step that predicts particle's positions and velocities; an \texttt{evaluation} step to evaluate new accelerations, their first order (\textit{jerk}), second order (\textit{snap}), and third order derivatives (\textit{crackle}); a \texttt{corrector} step that corrects the predicted positions and velocities using the results of the previous steps.

Code profiling shows the Hermite schema spends more than 90\% of time calculating the \texttt{evaluation} step, characterized by having an arithmetic intensity $I \simeq 10^{4}$ [FLOPs/byte] (ratio of FLOPS to the memory traffic) using $32^{3}$ particles.
In the following, time-to-solution and energy-to-solution measurements refer to that compute-bound kernel.

Three version of the code are available:

\begin{enumerate}[label=(\roman*)]
    \item \texttt{Standard C code:} cache-aware designed for CPUs and parallelized with hybrid MPI+OpenMP programming;
   
    \item \texttt{OpenCL code:} conceived to target accelerators like GPGPUs or embedded GPUs. All the stages of the Hermite integrator are performed on the OpenCL-compliant device(s). The kernel implementation exploits {\localm} memory (OpenCL terminology) of device(s), which is generally accepted as the best method to reduce global memory latency in discrete GPUs. However, on ARM embedded GPUs, the {\globalm} and {\localm} OpenCL address spaces are mapped to main host memory (as reported by the ARM developer guide\footnote{\url{https://bit.ly/2T1yrrw}}). So, a specific ARM-GPU-optimized version of all kernels of {\texttt{Hy-Nbody}}, in which {\localm} memory is not used, has been implemented and used in the results shown in the paper. The impact of such an optimization is shown in \cite{mynbody}.
    
    Regarding the host parallelization schema, a one-to-one correspondence between MPI processes and computational nodes is established and each MPI process manages all the OpenCL-compliant devices available per node (the number of such devices is user defined). Inside each share-memory computational node the parallelization is achieved by means of OpenMP.
    Such a implementation requires that particle data is communicated between the host and the device at each time-step, which gives rise to synchronization points between host and device(s). Accelerations and time-step computed by the device(s) are retrieved by the host on every computational node, reduced and then sent back again to the device(s);
   
    \item \texttt{Standard C targeting HLS tool:} Xilinx Vivado High Level Synthesis tool\footnote{\url{https://www.xilinx.com/products/design-tools/vivado/integration/esl-design.html}} was used to develop a highly optimized hardware accelerator for QFDB's FPGAs. The kernel was designed to be parameterizable, in order to experiment with different area vs performance implementations and to provide the capability of deploying it to any Xilinx FPGA with any amount of reconfigurable resources.
    
    Vivado HLS provides a directive-oriented style of programming where the tool transforms the high level code (C, C++, SystemC, OpenCL) to a Hardware Description Language (HDL) according to the directives provided by the programmer. Some of the optimizations performed in this kernel are described bellow:

    \begin{itemize}
       \item \textit{calculation in chunks:} Given the finite resources of the FPGA and the need to accelerate the Hermite algorithm in large arrays that exceed the amount of internal memory inside the FPGA (Block RAM), we followed a tiled approach where the kernel loops over the corresponding tiles of the original arrays and the core Hermite algorithm is performed in chunks of data stored internally. Block RAM is a low latency memory cell that can be configured in various widths and depths to store data inside the FPGA fabric but their capacity is limited (32.1 MB in our device). Thus, at the start of each computation the kernel fetches the data for the corresponding tile from main memory and stores it into Block RAM. During computation the partial results are also kept internally and as soon as the kernel finishes working with a tile, it writes the results back to the main memory and fetches the data for the next tile. Hence, the kernel has immediate access to the data it needs and communicates with the higher-latency DRAM only at the beginning and end of processing each tile, resulting in higher computational efficiency.
 
        \item \textit{burst memory mode:} As defined in the AXI4 protocol (which is used by the kernel to communicate with the DRAM) a "beat" is an individual transfer of a single data word, while a "burst" is a transaction in which multiple sequential data are transferred based upon a single address request. Since the data to be processed are stored sequentially in large arrays inside the DRAM, the kernel was implemented to request and fetch the data in bursts, and the burst size selected was the maximum burst size allowed by the AXI4, which is 4kB. This results in higher hardware complexity and resource utilization inside the FPGA fabric, but provides higher memory bandwidth and more efficient communication with the device's memory controller.

       \item \textit{loop pipeline, loop unroll, array partitioning:} Core Hermite algorithm has been pipelined achieving an initiation interval of 1 clock cycle. To achieve this, we increased the amount of the kernel's AXI4 read/write interfaces that communicate with the DRAM in order to fetch data from multiple arrays simultaneously. Also, by applying the loop unrolling directive we allowed the algorithm to be performed on more particles per cycle, with the corresponding increase of the FPGA resources needed due to the demand of more computational units. To achieve pipelining, the arrays stored internally were partitioned in multiple Block RAMs, since each BRAM has 2 ports for reading/writing and the kernel needs to access the data of many particles per cycle. These modifications minimized the idle time, considering the kernel is able to perform calculations on many particles in each individual cycle and remains idle only at the beginning and at the end of the processing of each tile when it communicates with the DRAM.
 
    \end{itemize}

     In our previous work \cite{mynbody2} we demonstrated a kernel showing a single QFDB's FPGA full potential. Due to its extra connectivity capabilities, the ''Network'' FPGA results in a higher reconfigurable resource congestion in order to operate. Thus, the above kernel's high demand of resources made it unfeasible to deploy it to the ''Network'' FPGA, so in order to demonstrate the application running in many FPGAs and split the computation load evenly inside the QFDB we chose a different size for this work's kernel. This kernel has 75\% throughput of the previous one and operates on a slightly higher frequency (320 MHz compared to 300 MHz). 
\end{enumerate}

\subsection{Floating point arithmetic considerations}
\label{subsection:floating}
Arithmetic precision plays a key role during the integration of the equations of motion of an $N$-body system. Generally, Hermite integration schema requires \texttt{double-precision} arithmetic in order to minimize the accumulation of the round-off errors, preserving both the total energy and the angular momentum during the simulation.
We have already demonstrated that \texttt{extended-precision} arithmetic \cite{ex} can speed up the calculation on GPUs, while is performance-poor on both CPUs and FPGAs \cite{mynbody2}, due to its higher arithmetic intensity compared to the \texttt{double-precision} algorithm (additional accumulations etc.).

Given that, we obtain the results shown in Section~\ref{section:results} using \texttt{double-precision} arithmetic to exploit both CPUs and FPGAs, while \texttt{extended-precision} arithmetic is employed to exploit GPUs.

\section{Computational performances and energy consumption}
\label{section:results}
\begin{figure}[h!]
    \centering
    \includegraphics[width=\textwidth]{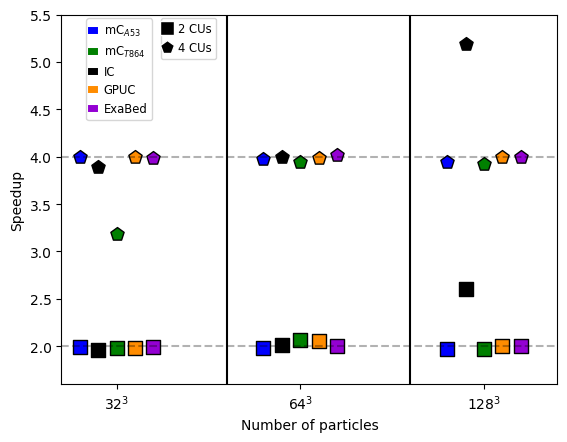}
    \caption{Speedup as a function of the number of particles using 2 and 4 {\CU}.}
    \label{fig:speedup}
\end{figure}

In all simulations, in the case of CPUs, the cores composing the {\CU} are exploited by means of OpenMP threads, and the multi-{\CU} by means of MPI; for GPUs, instead, we use a fixed number of 64 for the \texttt{work-group-size} (also called \texttt{block-size} in \texttt{CUDA} terminology)\footnote{We have already shown that the performance on ARM embedded GPUs is not driven by any specific \texttt{work-group-size}, regardless the usage of the local memory \cite{mynbody}.}.

We investigate the time-to-solution of {\ourcode} running two different test series. First, keeping the number of {\CU} constant, we increase the number of particles. We run three simulations with $32^{3}$, $64^{3}$ and $128^{3}$ particles. Then, keeping the number of particles constant, we vary the number of {\CU} used, from 1 to 4.

In Figure~\ref{fig:speedup}, we report the speedup (the ratio of the execution time using one \texttt{CU} to the time utilizing multiple {\CU}) using 2 and 4 {\CU}.
Different symbols refer to a different number of {\CU} (square for 2 {\CU} and pentagon for 4 {\CU}).
As expected, for almost all types of {\CU}, a linear speedup time execution reduction is achieved. We observe a speedup time execution reduction using 4 {\CU} for $32^{3}$ particles only in the case of the mC$_{\texttt{T864}}$. For the IC, on the contrary, a super linear speedup time execution is achieved using 2 and 4 {\CU} for $128^{3}$ particles.

\begin{figure}[h!]
    \centering
    \includegraphics[width=\textwidth]{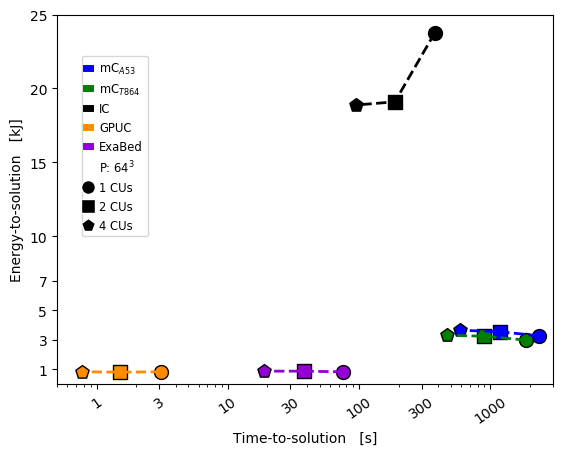}
    \caption{The performance-consumption plane for $64^{3}$ particles varying the number of {\CU}.}
    \label{fig:energy_time_to_solution}
\end{figure}

\begin{figure}[h!]
    \centering
    \includegraphics[width=\textwidth]{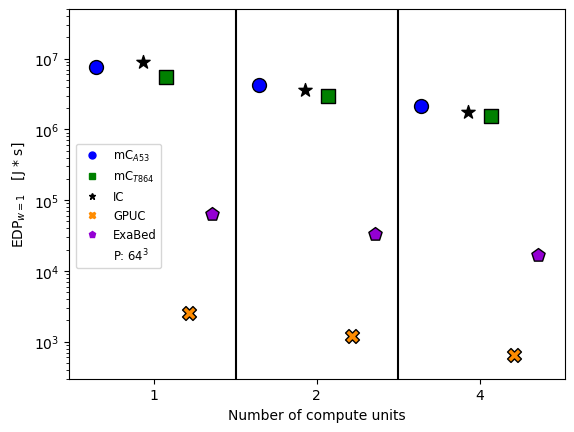}
    \caption{EDP as a function of the {\CU} using $64^{3}$ particles.}
    \label{fig:EDP}
\end{figure}

Figure~\ref{fig:energy_time_to_solution} shows the performance-consumption plane (energy-to-solution, {\ework}, vs time-to-solution) using $64^{3}$ particles and varying the number of {\CU}.

{\ourcode} is a compute-bound application, as stated in Section~\ref{section:code}, hence,
the latency of MPI communications across different {\CU} is negligible respect to the computational time. For this reason, in these tests we measure the computing performance of the {\CU} but not the network contribution.

Not surprisingly, both GPUC and ExaBed achieve the solution with less energy ({\ework}) than either IC or mC. The most interesting thing to point out is the equivalence of the energy-to-solution between GPUC and ExaBed, which indicates a definite trend toward Exascale prototype.
We can also see the effect of the energy consumption overhead when a node uses only a subset of its cores or sockets. This effect is evident for IC when using 1 or 2 {\CU} (a subset of all the {\CU} available).

In Figure~\ref{fig:EDP}, we present the results of the EDP for $w=1$ and $64^{3}$ particles.
We note that with the same {\CU}, the GPUC has a better EDP than the other platforms. When comparing the ExaBed and IC for the same time-to-solution configuration, the ExaBed has a better EDP. The configuration with 1 {\CU} on ExaBed has the same time-to-solution of the configuration with 4 {\CU} of the IC (we compare the execution time of the ExaBed using one \texttt{CU} with the execution time of the IC using four {\CU} on Figure~\ref{fig:energy_time_to_solution}).

\section{Conclusion and future work}
\label{section:conclusion}
In this work, we discuss the performance evaluation of four platforms concerning both the time-to-solution and energy-to-solution for code coming from the AA sector. 
Two platforms that represent the current status of HPC systems, the former Intel-based (IC) and the latter equipped with NVIDIA-Tesla-V100 GPUs (GPUC), an ARM MPSoC micro-cluster (mC) that could represent a low-budget HPC solution, and the ExaNeSt exascale prototype (ExaBed) that (possibly) represents the next generation of HPC systems.

Our analysis has been conducted using code for scientific production exploiting multi-CPUs, GPUs and FPGAs of the aforementioned platforms. 
The compute-bound nature of our application allows us to focus on performance assessment of the computational power and energy-efficiency of the devices, without dealing with the interplay of different key factors, like memory bandwidth, network latency and application execution pattern.

The overall picture, where accelerators outperform CPUs in terms of both performance and energy-efficiency, is not surprising. Exploiting  CPUs, when we set-up a run on the ExaBed in order to achieve the same time-to-solution with a run on the IC (ARM-A53 cores equip both the ExaBed and the mC), our results show that the former proves to be more power-efficient than the latter, which supports the exascale perspective of having single compute units to be tailored to a better FLOP/W ratio than pure FLOPs performance. 

Regarding accelerators, the high-end NVIDIA-Tesla-V100 GPUs perform faster than Xilinx US+ FPGAs on SoC, however the latter demonstrate superior energy-efficiency (the energy-to-solution is the same). We found that FPGA programming practice continues to be challenging for HPC software developers, even using the high-level-synthesis technique, which allows the conversion of an algorithm description in high level languages (e.g. C/C++, OpenCL) into a digital circuit. In comparison, GPU programming is pretty straightforward using the latest frameworks like CUDA, OpenCL or OpenAcc, but our great deal of effort has been devoted to optimize the kernel using \texttt{extended-precision} arithmetic. So at the end, we use comparable development effort in terms of design time and programmer training.

Our conclusion is that, when performance alone is a priority, CPUs or embedded GPUs on MPSoC are not a valid option, albeit their power-efficiency. ARM-based exascale prototypes may soon evolve to become a viable option for exascale-class HPC production machines if their performance improves while still maintaining a favorable power consumption.
Furthermore, in order to reduce the programming effort, the software environment should provide a clear, high-level, abstract interface to the programmer to efficiently execute functionality in the coupled-FPGAs, 
opening the path for successful and cost-effective use of such devices in HPC.

Our future activity will be aimed to exploit more computational nodes, offering a more comprehensive benchmark of both the computation power and the interconnect network of the platforms.

\section{Acknowledgments}

This work was carried out within the EuroExa FET-HPC and ESCAPE projects (grant no. 754337 and no. 824064), funded by the European Union's Horizon 2020 research and innovation program. 
We thank the INAF Trieste Astronomical Observatory Information Technology Framework.
We thank Piero Vicini and the INFN APE Roma Group for the support and for the use of INFN computational infrastructure.
We also thank  Giuseppe Murante and Stefano Borgani for the fruitful discussions on the energy and performance optimization of our codes.
This research has been made use of IPython \cite{IPython}, Scipy \cite{SciPy}, Numpy \cite{NumPy} and MatPlotLib \cite{MatPlotLib}.


\begin{thebibliography}{99}


\bibitem{Calore_2015}
Calore E., Schifano S. F., Tripiccione R.: Energy-Performance Tradeoffs for HPC Applications on Low Power Processors. Euro-Par 2015: Parallel Processing Workshops. Springer International Publishing (2015). doi: 10.1007/978-3-319-27308-2\_59

\bibitem{Nikolskiy_2016}
V. P. Nikolskiy, V. V. Stegailov and V. S. Vecher: Efficiency of the Tegra K1 and X1 systems-on-chip for classical molecular dynamics. (2016) International Conference on High Performance Computing \& Simulation (HPCS), Innsbruck, 2016, pp. 682-689.

\bibitem{Nikolskiy_2018}
Nikolskii V., and Stegailov, V.: Domain-Decomposition Parallelization for Molecular Dynamics Algorithm with Short-Ranged Potentials on Epiphany Architecture. Lobachevskii Journal of Mathematics (2018). doi:10.1134/S1995080218090159

\bibitem{Morganti_2016}
Morganti L, Cesini D, Ferraro A: Evaluating Systems on Chip through HPC Bioinformatics and Astrophysics Applications. 24th Euromicro International Conference on Parallel, Distributed, and Network-Based Processing (PDP) 2016, 541-544, 2016. doi: 10.1109/PDP.2016.82









\bibitem{exanest2}
Ammendola R., Biagioni A. , Cretaro P.,  Frezza O., Cicero FL et al.: The Next Generation of Exascale-Class Systems: The ExaNeSt Project. In Euromicro Conference on Digital System Design (DSD), Vienna,  pp. 510-515 (2017) \url{http://dx.doi.org/10.1109/DSD.2017.20}





\bibitem{Taffoni_ARM} Taffoni, G.; Bertocco, S.; Coretti, I.; Goz, D.; Ragagnin, A.; Tornatore, L.. "Low Power High Performance Computing on Arm System-on-Chip in Astrophysics" doi: 10.1007/978-3-030-32520-6\_33


\bibitem{bertocco} Bertocco, S.; Goz, D.; Tornatore, L.; et al. ``INAF Trieste Astronomical Observatory Information Technology Framework''
doi: arXiv:1912.05340 [astro-ph.IM]
4 pages, conference, ADASS 2019

\bibitem{taffoni_chipp} Giuliano, Taffoni; Ugo, Becciani; Bianca, Garilli; et al. ``CHIPP: INAF pilot project for HTC, HPC and HPDA'' 
doi:arXiv:2002.01283 [astro-ph.IM]
4 pages, conference, ADASS 2019

\bibitem{czarnul} Pawel, Czarnul; Jerzy, Proficz; and Adam Krzywaniak, ``Energy-Aware High-Performance Computing: Survey of State-of-the-Art Tools, Techniques, and Environments' Scientific Programming 2019, Article ID:8348791, Hindawi

\bibitem{dutot} P. Dutot, Y. Georgiou, D. Glesser, L. Lefevre, M. Poquet and I. Rais, "Towards Energy Budget Control in HPC," 2017 17th IEEE/ACM International Symposium on Cluster, Cloud and Grid Computing (CCGRID), Madrid, 2017, pp. 381-390.

\bibitem{cesiniCosa} Daniele, Cesini; Elena, Corni; Antonio, Falabella; and et al., ``Power-Efficient Computing: Experiences from the COSA Project'' Scientific Programming 2017, Article ID:7206595, Hindawi

\bibitem{Power_DPSNN} Simula, F. et al., ``Real-Time Cortical Simulations: Energy and Interconnect Scaling on Distributed Systems'' 2019 27$^{th}$ Euromicro International Conference on Parallel, Distributed and Network-Based Processing (PDP), Pavia, Italy, 2019, pp. 283-290.

\bibitem{Brain_LowPower} Ammendola, R.; Biagioni, A; Capuani F.; Cretaro, P.; De Bonis, G.; Lo Cicero, F.; Lonardo, A.; Martinelli, M.; Paolucci, P.S.; Pastorelli, E.; Pontisso, L.; Simula, F; and Vicini, P., ``The Brain on Low Power Architectures - Efficient Simulation of Cortical Slow Waves and Asynchronous States'', ParCo 2017, IOS Press, Advances in Parallel Computing Volume 32: Parallel Computing is Everywhere, pp. 760-769. doi:10.3233/978-1-61499-843-3-760

\bibitem{taffoni} Taffoni, Giuliano; Murante, Giuseppe; Tornatore, Luca; Katevenis, Manolis; Chrysos, Nikolaos; Marazakis, Manolis: Shall Numerical Astrophysics Step Into the Era of Exascale Computing?, 2019ASPC..521..567T

\bibitem{ska} P. E. Dewdney,  P. J. Hall, R. T. Schilizzi,  and T. J. L. Lazio, ``The Square Kilometre Array'', IEEE Proceedings, 97, 1482, August 2009

\bibitem{cta} B. S. Acharya, M. Actis, T. Aghajani, and et al., ``Introducing the CTA concept'', in  Astroparticle Physics, 43, 3, March 2013

\bibitem{elt} T. de Zeeuw, R. Tamai, and J. Liske, ``Constructing the E-ELT'', The Messenger, 158, 3, Dicember 2014
%
\bibitem{jwst} J.P. Gardner, J.C. Mather, M. Clampin, and et al. ``The James Webb Space Telescope'',  Space Sci.Rev., 123, 485. April 2006
%
\bibitem{euclid} L. Amendola, S. Appleby, D. Bacon, T. Baker, M. Baldi, N. Bartolo, and et al., ``Cosmology and Fundamental Physics with the Euclid Satellite,'' Living Reviews in Relativity, 16:6, September 2013
%
\bibitem{erosita} A. Kolodzig, M. Gilfanov, R. Sunyaev, S. Sazonov, and M. Brusa. AGN and QSOs in the eROSITA All-Sky Survey. I. Statistical properties. A\&A, 558:A89, October 2013.

\bibitem{euroexa}  EuroEXA: European Exascale System Interconnect and Storage. \url{https://euroexa.eu/}

\bibitem{higpus_1} Spera M., Capuzzo-Dolcetta R., "Rapid mass segregation in small stellar cluster"s, 2017Ap\&SS.362..233S, doi:10.1007/s10509-017-3209-6

\bibitem{higpus_2} Spera, M., Mapelli, M., Bressan, A.,  "The mass spectrum of compact rem-nants  from  the  PARSEC  stellar  evolution  tracks". 2015MNRAS.451.4086S, doi: 10.1093/mnras/stv1161

\bibitem{KATEVENIS201858}M. Katevenis, R. Ammendola, A. Biagioni, P. Cretaro, O. Frezza, F. Lo Cicero, and et al., ``Next generation of Exascale-class systems: ExaNeSt project and the status of its interconnect and storage development,'' Microprocessors and Microsystems, 61, 58, 2018

\bibitem{exanest} Katevenis M., Chrysos N., Marazakis M.,  Mavroidis I., Chaix F., Kallimanis N., et al.: The ExaNeSt Project: Interconnects, Storage, and Packaging for Exascale Systems, 2016 Euromicro Conference on Digital System Design (DSD), Limassol, pp. 60-67 (2016)

\bibitem{qfdb} F. Chaix, A.D. Ioannou, N. Kossifidis, N. Dimou, G. Ieronymakis, M. Marazakis, V. Papaefstathiou, V. Flouris, M. Ligerakis, G. Ailamakis, T.C. Vavouris, A. Damianakis, M. G.H. Katevenis and I. Mavroidis, "Implementation and impact of an ultra-compact multi-FPGA board for large system prototyping", 5th International Workshop on Heterogeneous High-performance Reconfigurable Computing ($H^2RC'19$), held in conjunction with SC'19, 2019

\bibitem{incass} S. Bertocco, D. Goz, L. Tornatore, G. Taffoni, ``
INCAS: INtensive Clustered ARM SoC -- Cluster Deployment,'' in INAF Technical Reports doi:10.20371/INAF/PUB/2018\_0000, 2018

\bibitem{slurm} J.A. Pascual, J. Navaridas, J. Miguel-Alonso, ``Effects of Topology-Aware Allocation Policies on Scheduling Performance. Job Scheduling Strategies for Parallel Processing,'' Lecture Notes in Computer Science. 5798, pp. 138--144, 2009

\bibitem{edp}
Cameron K.W., Ge R., Feng X., Varner D., Jones C.: High-performance, power-aware distributed computing framework. In Proceedings of the International Conference on High Performance Computing, Networking, Storage, and Analysis (SC), ACM/IEEE, (2004)

\bibitem{mynbody} Goz, D., Bertocco, S., Tornatore, L., and Taffoni, G. ``,Direct N-body Code on Low-Power Embedded ARM GPUs,'' Intelligent Computing, Springer International Publishing, Charm, pp. 179--193, 2019

\bibitem{mynbody2} Goz, D. Ieronymakis, G. Papaefstathiou, V. Dimou, N. Bertocco, S., Ragagnin, A. Tornatore, L. Taffoni, G. and Coretti, I., "Direct N-body application on low-power and energy-efficient parallel architectures", 2019, arXiv, arXiv:1910.14496

\bibitem{nbody:4}
Capuzzo-Dolcetta, R., Spera, M., Punzo, D.: A fully parallel, high precision, N-body code running on hybrid computing platforms. Journal of
Computational Physics 236 (2013) 580–593

\bibitem{nbody:5}
Capuzzo-Dolcetta R., Spera M.: A performance comparison of different graphics processing units running direct N-body simulations. Computer Physics Communications 184:2528–2539 (2013)

\bibitem{nbody:6}
Spera M.: Using Graphics Processing Units to solve the classical N-body problem in physics and astrophysics. ArXiv e-prints 1411.5234 (2014)

\bibitem{hermite}
Nitadori K., Makino J.: Sixth- and eighth-order Hermite integrator for N-body simulations. New Astronomy 13:498–507, (2008)

\bibitem{ex}
Thall A.: Extended-precision floating-point numbers for gpu computation. p 52, (2006)  DOI 10.1145/1179622.1179682

\bibitem{NumPy} van der Walt, S. and Colbert, S.C. and Varoquaux, G., doi:10.1109/MCSE.2011.37

\bibitem{IPython} Perez, F. and Granger, B.E, doi:10.1109/MCSE.2007.53

\bibitem{SciPy} SciPy, "SciPy: Open source scientific tools for Python, url:"http://www.scipy.org/", doi:https://doi.org/10.1038/s41592-019-0686-2

\bibitem{MatPlotLib} Hunter, J.D., doi:10.1109/MCSE.2007.55

\end{thebibliography}
\end{document}